\documentclass[preprint,aps,nofootinbib]{revtex4}
\usepackage{graphicx}
\usepackage{epstopdf}
\usepackage{subfigure}
\usepackage{textcomp}
\def\bkR{{\rm I\kern-.17em R}}
\def\bkC{{\rm \kern.24em \vrule width.05em height1.4ex depth-.05ex \kern-.26em C}}

\def\be{\beta}

\def\frac#1#2{{\textstyle{{#1}\over {#2}}}}

\def\lsim{\mathrel{\rlap{\lower4pt\hbox{\hskip1pt$\sim$}}
    \raise1pt\hbox{$<$}}}
\def\gsim{\mathrel{\rlap{\lower4pt\hbox{\hskip1pt$\sim$}}
    \raise1pt\hbox{$>$}}}
\def\sqr#1#2{{\vcenter{\vbox{\hrule height.#2pt
         \hbox{\vrule width.#2pt height#1pt \kern#1pt
         \vrule width.#2pt}
         \hrule height.#2pt}}}}

\def\laq{\raise 0.4 ex \hbox{$<$}\kern -0.8 em\lower 0.62 ex\hbox{$\sim$}}
\def\gaq{\raise 0.4 ex \hbox{$>$}\kern -0.7 em\lower 0.62 ex\hbox{$\sim$}}

\def\be{\begin{equation}}
\def\ee{\end{equation}}
\def\ba{\begin{eqnarray}}
\def\ea{\end{eqnarray}}

\def\dalemb#1#2{{\vbox{\hrule height.#2pt
        \hbox{\vrule width.#2pt height#1pt \kern#1pt \vrule width.#2pt}
        \hrule height.#2pt}}}

\def\dalemb#1#2{{\vbox{\hrule height.#2pt
        \hbox{\vrule width.#2pt height#1pt \kern#1pt \vrule width.#2pt}
        \hrule height.#2pt}}}

\def\gtorder{\mathrel{\raise.3ex\hbox{$>$}\mkern-14mu
             \lower0.6ex\hbox{$\sim$}}}
\def\ltorder{\mathrel{\raise.3ex\hbox{$<$}\mkern-14mu
             \lower0.6ex\hbox{$\sim$}}}

\begin{document}

\rightline{October 2011}

\title{Digging down the past}

\author{Orfeu Bertolami\footnote{Also at Instituto de Plasmas e Fus\~ao Nuclear,
 Instituto Superior T\'ecnico, 
Avenida Rovisco Pais 1, 1049-001 Lisboa, Portuga. E-mail: orfeu.bertolami@fc.up.pt}}

\vskip 0.3cm

\affiliation{Departamento de F\'\i sica e Astronomia  \\
Faculdade de Ci\^encias, Universidade do Porto \\ 
Rua do Campo Alegre 687, 4169-007 Porto, Portugal
}


\vskip 3cm

\begin{abstract}

\vskip 1cm

{It is often stated that cosmology is an extreme form of archeology that deals with aeons of time extending back to thousands of millions of years. One talks, for instance, of relics of the early Universe, and phases in the history of the Universe 
that resemble the division of phases in human development used by archeologists to interpret the data they gather in the field. In this text we shall analyze these ``linguistic" similarities, and the differences between these two scientific activities 
paying some attention on the role played by their theoretical guidelines.
}

\end{abstract}

\maketitle

\section{Introduction}


Cosmology is regarded within physics as a radical form of ``archeology" as it aims to understand events that took place thousands of millions of years ago. One often use terms like {\it relics} of the early Universe, and well defined periods in the evolution of the Universe that resemble the historical divisions which characterize the evolution of humankind. In a broad sense, these linguistic similarities, indicate that there are some common features between cosmology and archeology; however, there are striking differences of scope, methods and elements of analysis, and as we shall see, in the role played by theory in these two scientific activities. In what follows we shall examine these differences and speculate on a possible dialogue between cosmology and archeology. Given the obvious limitations of this author competences, more attention will be paid on describing cosmology and the picture it yields of the cosmos and its evolution since the Big Bang. The reason for this is clear: this author feels that this text can be much more useful if it aims to provide some guidance and insight on the most recent developments in cosmology to the more human sciences oriented readers, than to pretend to show some erudition or to master a subject that is not his one.     

In the past decades, cosmology has acquired the distinct features of an activity driven and supported by observations. This is a most welcome development in a subject that sprang from purely theoretical reasoning in the seminal work of Einstein in 1917 \cite{Einstein1917} in the context of his 1915's theory of general relativity.  Subsequent theoretical developments by Friedmann and Lem\^aitre in the 1920's and Gamow and collaborators in the 1950's led to the idea of an evolving Universe, whose most striking observational implication was the momentous discovery by Hubble in 1929 that the Universe is expanding.  Decades of development led eventually to the emergence of a robust framework for the understanding of all events in the Universe, the Hot Big Bang (HBB) model. In the context of this model, it is possible to infer that the Universe has arisen from a space-time explosion that took place about 13.7 thousand millions years ago (see Ref. \cite{Bertolami2006a} for a thorough discussion).

The HBB model allows for highly non-trivial predictions and makes sense of an enormous body of evidence in astronomy, astrophysics and observational cosmology. The HBB is thus a fully operational physical model in the strictest  sense of the term, and can be regarded as a ``paradigm" whose corner stones are the general theory of relativity, quantum field theory, nuclear physics, statistical mechanics, {\it etc}. In fact, predictions and computations within the HBB use virtually all known areas of physics, from high-energy physics to nuclear and atomic physics, from statistical mechanics to plasma physics. 
The HBB model allows for a detailed description of the Universe evolution, and this reconstruction includes 
fairly complex and highly non-linear phenomena such as the formation of structures like galaxies, clusters of galaxies and superclusters of galaxies. Indeed, within the so-called Cold Dark 
Matter\footnote{It is believed that dark matter is responsible for the cohesion of galaxies and clusters of galaxies, for the observed bending of light in the so-called gravitational lenses, among other phenomena (see {\it e.g.}  Refs. \cite{Einasto,Roos})} model \cite{Bertolami2006a}, the structure formation scenario triggered by non-relativistic particles (``cold" particles)
after the decoupling from the other Universe's constituents, one can, through N-body computer simulations driven by the law of gravity, to
match the observed large scale matter distribution in the Universe and predict the size of the 
smallest structural seeds, actually  
Earth-mass structures \cite{Diemand}. More 
recently, it is even possible to get some knowledge about formation of the very first stars \cite{Roseta}.

Furthermore, the HBB model, matches impressively well the observed abundance of light elements, $He^4$, $He^3$, $D$ and $Li^7$,  
which, according to the model, were 
synthesized a few minutes after the Big Bang, 
when temperatures were about $10^{11~o}$K. 
This is a well defined prediction of the model that clearly allows for its falsification. This prediction implies the remaining elements had to be synthesized 
in the interiors of the stars, and the observation of early stars with rather few 
elements is yet another consistency check of the model.   

Of course, one could conceive different cosmological scenarios based on some alternative of theory of gravity instead of general 
relativity. However, the fundamental underlying principle of
general relativity, namely the connection between space-time curvature and matter-energy as established by Einstein's 
field equations, is consistent with all experimental evidence to considerable accuracy (see e.g. Refs. \cite{Will05,BPT06} for
reviews). Despite that, there are a number
of reasons, theoretical and experimental, to question the general theory of relativity as the
ultimate description of gravity (see e.g. \cite{Bertolami2006b,OBP2006} and references therein). 
These concerns are related with fundamental issues, such as 
the singularity problem, the cosmological constant problem and the underlying mechanism of inflation, 
difficulties that cannot be satisfactorily addressed within the context of general relativity \cite{Bertolami2006a}. 
Therefore, it is not impossible that the Big Bang scenario based as it is on general relativity, may turn out 
to be a fundamentally incomplete description of the space-time evolution. Even though, experimental evidence strongly suggests that it is 
not at all the case.
 
Actually, cosmology is the natural testing ground for unification ideas, given that the description of the Universe's history and 
evolution requires, as mentioned, the integration of all physics. In fact, the HBB model allows for  
the interpretation of all known observational facts within a coherent picture, if one admits the existence of states beyond the {\it Standard Model} of the strong, weak and electromagnetic interactions:  {\it dark energy}, which in 
its simplest form can be just a (fairly small) cosmological constant in Einstein's field equations, and {\it dark matter}. The HBB requires, at very early times  ($t \simeq 10^{-35}$ seconds), 
a period of accelerated expansion dubbed {\it inflation}.
Inflation explains the features of the cosmic microwave background radiation (CMBR) in a causal way and solves the so-called horizon, homogeneity and rotation problems. Inflation also suggests 
an elegant mechanism for structure formation, based on the ubiquitous quantum fluctuations that all fields are subjected to, 
and in particular, the presumed scalar field that accounts for inflation, the {\it inflaton}. In the context of {\it Grand Unified Theories} (GUT) (see {\it e.g.} Ref. \cite{Ross} for an extensive discussion), 
inflation can prevent the dominance of magnetic monopoles, which would lead to the Universe's collapse just after the Big Bang. Inflation connects, in a direct way, cosmology with GUTs, supergravity, superstrings, 
that is with the physics of the very early Universe  
(see {\it e.g.} Ref. \cite{Bertolami1987}).     
 
The discovery in 1998 of the late acceleration of the Universe \cite{Perlmutter} as inferred from the observation of far away type Ia supernovae, has lead to the need of considering the existence of {\it dark energy}, 
and likewise inflation, of an hypothetical scalar field, dubbed {\it quintessence}, to drive the accelerated expansion. The putative unification of dark energy and 
dark matter \cite{Bento2002} is one the ideas that can be discussed in the context of the HBB model, with quite interesting observational implications (see {\it e.g.} Ref. \cite{Barreiro2008}). 

Moreover, from the observation of the features of the CMBR, the surface of last 
interaction between matter and radiation corresponding to the transition that allowed the radiation to freely propagate, one can estimate the age of the Universe as to be about 13.7 thousand million years (Gys). Actually, the CMBR is a picture of the Universe at about 370 thousand years after the Big Bang. Furthermore, the so-called Hubble ultra deep field observations allow one to estimate that the first proto-galaxies were formed more than
12 Gys ago. To this cosmological chronology of events one could add the additional cornerstones: 
radiative dating allows for estimating the age of Earth as being about 
4.5 Gys; evidence arising from dating of stromatolites fossils suggests 
that life appeared on Earth 3.8 Gys ago; the first macroscopic fossils 
seem to have first shown up 700 million years ago; the tectonic 
processes that gave origin to the Atlantic ocean took place about 100 million 
years ago and our primate ancestors walked on their inferior members about 
3 million years ago; the first human population's organizations appeared 
40 thousand years ago and the impressive first human artistic 
manifestations, the cave and rock paintings found in Europe and Australia, date 
back 20 to 30 thousand years. Thus, the HBB model allows one to make sense of a consistent timeline of events in the history of the Universe.

\section{{\bf Cosmology $\times$ Archeology}}

On quite general terms, we could say that both, cosmology and archeology, are sciences based on evidence obtained from digging down the past, on surveying for relics, in order to create a consistent scenario of events . This apparent similarity is however, not quite exact, as cosmology is, on its very foundations, an essentially theoretically driven science, somewhat in opposition to archeology (see below). Actually, in the past, some authors did question the scientific standing of cosmology, but recent observational developments, as discussed above, have fundamentally changed this perspective, even though, for some, cosmology is no quite like other physical theories \cite{Goenner} as, strictly speaking, factual evidence is not directly accessible to experimental confirmation, but only to observational verification, likewise in astronomy and astrophysics.   
 
These considerations imply that ``relics" in cosmology are identified through laboratory instruments that observe a specific physical feature whose origin is in the past, 
such as the light, or more generally the electromagnetic radiation, from a far away astrophysical or cosmological sources. This means that, in opposition to archeology, 
whose relics correspond to artifacts, utensils, architecture and cultural landscapes that can be recognized by their practical utility and their ethnocentric and cultural role, in cosmology, those relics are past events that manifest themselves as physical phenomena with a clear counterpart in the laboratory. So, it seems just natural that one expects crucial information for cosmology arising from the Large Hadron Collider at CERN, from the laboratory searches for dark matter, from the neutrino SuperKamiokande-III experiment, from the Planck Surveyor of the CMBR, from the GLAST/Fermi gamma radiation experiment, {\it etc}.

Another common issue in cosmology and archeology is the central role played by the process of dating. This procedure requires necessarily a correlation with a known physical system whose intrinsic time scale is known. Radiative dating, a technique widely used in various branches of science, archeology, paleontology, geology, etc, turns out to be also useful to correlate with  the greatest cosmological aeons. This implies that cosmology has its own time scale to correlate with the greatest radiative time scales, which involve knowledge, for instance, of the ratio of life times of $U^{235}/U^{238}$ and $Th^{232}/U^{238}$, that allow dating processes with time scales ranging from $6.6 \times 10^9$ years to about $10\times 10^9$ years. These figures are comparable with the typical time scale of cosmology, which corresponds to the inverse of the expansion rate of the Universe, Hubble's constant, $H_0 \simeq 70~km/(sMpc)$, that is about $7 \times 10^9$ years. The fact that the Universe is somewhat older than $H_0^{-1}$ reflects the fact that dark energy has played a relevant role in the most recent dynamics of the Universe. 

In this context, one can ask about the greatest and smallest time and length scales in cosmology and archeology. In cosmology, these scales are obtained from physical arguments, while in archeology these scales are essentially given by field surveying constraints . Indeed, the greatest   
time scales in cosmology can be associated to the horizon scale, the length scale which sets the ``size" of the Universe as a result of its expansion history, being typically some orders magnitude of the bench mark value $10^{26}~km$, the observable horizon size, the age of the Universe times the velocity of light, approximately $300000~km/s$. The smallest length scale is the so-called Planck length, which corresponds to the scale where strong gravitational fields are affected by quantum mechanics effects, being of order of the incredibly small length scale, namely $10^{-35}~m$. In archeology, the large time scales is presumably several decades of thousands of years and the smallest time scale corresponds to the smallest time interval that can be inferred from gathered evidence, which typically is of order of few decades. Of course, the greatest scale of archeology can be considerably extended if one considers the dawn of human history from the development of the first stone tools in eastern Africa, about 3.4 million years ago. Actually, over 99$\%$ of human history took place in prehistoric societies, of which no written records exist, from the Palaeolithic (from about 2.6 millions years ago to about 10000 yeas ago) until the advent of literacy, about 6000 years ago \cite{Cook}. 

Of course, the span of these time scales, from the smallest to the greatest, are based in cosmology on the continuity of the physical world, while in archeology this connection arises from the continuity of the human existence, which in practice corresponds to the rise and fall of human activities, civilizations, etc.  But, besides this disparity of scales, it is obvious that cosmology and archeology dig down for different forms of evidence from the past. Cosmology aims to reconstruct the history of space-time from its very beginning, describing the behaviour of matter in it. On its hand, the ultimate goal of archeology is to trace human existence from its very origin till the recent past, meaning that methodologically, archeology lies necessarily within a broader class of human sciences, that includes history, ethnology, physical anthropology, geography, etc, as well as linguistics, chemistry and branches of biology such as paleontology, paleozoology, paleobotany and the genetics of populations.

\section{{\bf The role of the theory}}

As already mentioned, cosmology can only be thought through a theory of global space-time, general relativity. The evolving picture of the Universe arises as solutions of Einstein's field equations under certain assumptions about the large scale structure of space-time. The validity of a given set of solutions, and of the assumptions they are based upon, is assessed by their capability to match the observed evolution of the formed structures in a space-time that is expanding and that, at late times, is accelerating. But it is important to emphasize that without a theoretical framework, it is virtually impossible to make sense of the data gathered in observational cosmology. Without the HBB model, the features of the CMBR, for instance, would be no more than a meaningless set of temperature readings.  

Of course, one is assuming that the validity of the theory itself is established independently of cosmological considerations. Actually, a vivid discussion is currently under way, whether the standard description of the Universe dynamics based on general relativity together with dark energy and dark matter can be equally well accounted by some alternative theory of gravity without these dark components \cite{Bertolami2008a,BertolamiParamos2010,BFP2010}. But, the main point here is that cosmology makes little sense without a theoretical framework to understand the global structure of space-time and to account data in a way that is compatible with causality, a fundamental ingredient of any physical theory  \cite{Bertolami2008b,BertolamiLobo2009}. It remains to be seen whether general relativity is the ultimate theory to achieve this description, even though it is widely believed that general relativity is, however impressive, just a provisional step towards  the complete description.   

In archeology, it seems that, likewise in all scientific endeavour, but most particularly in social sciences, theoretical thinking is very much focussed on the way data should be gathered, regarded and interpreted. That is to say, that the theory sets the methodological procedure on how to survey, how to gather, which, in archeology, most often involves excavation, how to interpret and to link the data. This suggests that archeologists can interpret their data without a unique theoretical framework (see Ref. \cite{Greene} and texts there attached). 

Indeed, from its origins in the late 19th century, it was assumed that the goal of archeology was to explain why human groups changed and adapted, with special  emphasis on the historical particularism of each culture.  In the early 20th century, most archaeologists viewed past societies as if they were directly and continuously connected to existing ones \cite{Greene}. However, in the 1960s, an archaeological movement arose that rebelled against the established cultural-historic archaeology and proposed an approach based  on more  anthropological methods, with stress on the testing of hypotheses and on the scientific method, a movement that became known as {\it processual archaeology}. There followed in the 1980s, a post-modernist inspired movement, and whose main features were the questioning of processualism's claims of scientific positivism and impartiality, and that asked for a more self-critical theoretical approach. Naturally, this movement has been criticized by processualists on the grounds of its lack of scientific rigor. The debate between the validity of processualism and post-processualism approaches is still on, but meanwhile, a synthesis is emerging,  the {\it historical processualism} approach, with a wide range of influences, from neo-Darwinian evolutionary thinking to cognitive science, and  gender-based and Feminist oriented archaeology studies.

 \section{Cosmology and Archeology: a possible dialogue? }
 
After discussing the common features and the most salient differences between cosmology and archeology, one may wonder whether a dialogue between these disciplines is possible. In some recent reflections, this author has discussed about the cultural and philosophical implications of the most recent developments in cosmology \cite{Bertolami2008c, Bertolami2010b}, with the conclusion that it is just natural to expect that a deeper grasping of our standing in the Universe does have an impact on the way we regard the human condition. If so, we can say that cosmology does influence other branches of scientific activity, and in particular, archeology, given the common features discussed above. 

Moreover, the reductionist approach of modern science, whose main feature is the description of complex phenomena in the world through the understanding of the dynamics and interaction of the smallest and invariant physical entities, such as elementary particles\footnote{The concept of elementary particle can be regard in the absolute sense, when it refers to the smallest identifiable particle, such as {\it quarks}, {\it leptons} and the {\it scalar} and {\it vector bosons} of the four fundamental interactions of nature. These are the {\it gluons} for the strong nuclear interactions, the {\it photon}, {\it bosons} $W^{\pm}$ and $Z^0$ and the Higgs boson (if it is detected) for the electroweak interaction, and the {\it graviton} for the gravitational interaction. But, the concept can be also considered in a relative sense when it refers to the smallest identifiable structure at a given scale of energy, such as for instance, the atom in atomic physics, the {\it nucleons} (protons and neutrons) in nuclear physics, etc.} naturally leads to the idea of encompassing the interactions of the parts in a conceptually simpler and unified scheme. Actually, as discussed elsewhere \cite{Bertolami2010a}, the ``unification principle" has already yielded remarkable results, being thus, a rather general methodological procedure, at least in physics. 

Recent developments towards a synthesis of the various theoretical and methodological approaches in social sciences, and actually also in biology \cite{Smocovitis}, indicate a mutual influence that is, in the opinion of this author, beneficial for all sciences.  

In fact, our current ``Weltanschauung'' is based on an impressive body of experimental and observational facts, 
and theoretical reasoning, which is on its own, an unified approach encompassing various branches of human knowledge such as physics, astronomy, chemistry, geology, paleontology, biology, genetics, archeology, history, etc. This unified view yields a most impressive description of the Universe's evolution and of the human development. This picture allows for the understanding of the articulation of ``origins'', 
from the origin of the Universe to the emergence of life on Earth and its evolution till the first human social organizations. 
The development of the ideas and the gathering of observational facts that support this encompassing vision of the cosmos and of the human existence on Earth  
is one of the finest constructions of the human spirit. It is most pleasing to think that this construction is essentially open, and that it connects, in a perpetual way, the work of scores of scientists and thinkers from the past and present to those of the future generations.

\subsection*{Acknowledgments}

\vspace{0.3cm}

\noindent
I  would like to thank Professor V\'\i tor Jorge for the kind and provocative invitation to write this article and to Mariana Ara\'ujo, Maria Jos\'e Barbosa and F\'atima Mota for 
the critical reading of the manuscript.


\end{document}